\documentclass[twocolumn,showpacs,aip,superscriptaddress,reprint,floatfix]{revtex4-2}

\usepackage{bm}
\usepackage{xcolor}
\usepackage{graphicx}
\usepackage{lineno}
\usepackage{comment}
\usepackage{amsmath}
\usepackage{amssymb}
\usepackage{dsfont}
\usepackage{physics}
\usepackage[caption=false]{subfig} 
\usepackage{hyperref}
\usepackage{url}
\usepackage{amsmath,mleftright,mathtools}
\usepackage{wrapfig}
\usepackage[normalem]{ulem}

\usepackage[utf8]{inputenc}
\usepackage[T1]{fontenc}
\newcommand{\RED}[1]{\textcolor{black}{#1}}

\makeatletter
\def\@email#1#2{%
 \endgroup
 \patchcmd{\titleblock@produce}
  {\frontmatter@RRAPformat}
  {\frontmatter@RRAPformat{\produce@RRAP{*#1\href{mailto:#2}{#2}}}\frontmatter@RRAPformat}
  {}{}
}%
\makeatother

\begin{document}

\newcommand{\tf}{\textbf}
\newcommand{\bo}{\mathbf}
\newcommand{\br}{{\bf r}}
\newcommand{\be}{\begin{equation}}
\newcommand{\ee}{\end{equation}}
\newcommand{\bea}{\begin{eqnarray}}
\newcommand{\eea}{\end{eqnarray}}
\newcommand{\nup}{n_{\uparrow}}
\newcommand{\ndown}{n_{\downarrow}}
\newcommand{\Id}[1] {\int \! \! {\rm d}^3 #1}
\renewcommand{\v}[1]{{\bf #1}}

\def\qw		{{\bf q},\omega}
\def\ql  	{{\bf q},\lambda}
\def\rw		{{\bf r},\omega}
\def\rt		{{\bf r},t}
\def\rtp	{{\bf r},t'}
\def\tone   {t_{1}}
\def\ttwo   {t_{2}}
\def\taoone   {\tau_{1}}
\def\taotwo   {\tau_{2}}
\def\ttmto   {\tau_{2}-\tau_{1}}
\def\tomtt   {\tau_{1}-\tau_{2}}
\def\pp		{{\bf p}}
\def\rr		{{\bf r}}
\def\rrp	{{\bf r}'}
\def\rrone	{{\bf r}_1}
\def\rrtwo	{{\bf r}_2}
\def\vv		{{\bf v}}
\def\BB		{{\bf B}}
\def\QQ		{{\bf Q}}
\def\GG		{{\bf G}}
\def\RRI        {{\bf R}_{Is}}
\def\RRJ        {{\bf R}_{Js'}}
\def\RR		{{\bf R}}
\def\KK		{{\bf K}}
\def\EE		{{\bf E}}
\def\qq		{{\bf q}}
\def\kk		{{\bf k}}
\def\jj		{{\bf j}}
\def\calGzero   {{\cal G}^{(0)}(\kk,\gon)}
\def\calG       {{\cal G}(\kk,\gon)}
\def\intr	{\int\,d{\bf r}\,}
\def\intrp	{\int\,d{\bf r}'\,}
\def\introne	{\int\,d{\bf r}_{1}\,}
\def\intrtwo	{\int\,d{\bf r}_{2}\,}
\def\intao     {\int\,d\tau\,}

\def\ga         {\alpha}
\def\gb         {\beta}
\def\gc         {\gamma}
\def\gC         {\Gamma}
\def\gd         {\delta}
\def\gD         {\Delta}
\def\gee        {\epsilon}
\def\gl         {\lambda}
\def\gL         {\Lambda}
\def\go         {\omega}
\def\gO         {\Omega}
\def\goql       {\omega_{\qq \gl}}
\def\goqlt       {\omega_{\tilde \qq \gl}}
\def\gql        {\qq \gl}
\def\gon        {\omega_{n}}
\def\goj        {\omega_{j}}
\def\goi        {\omega_{i}}
\def\igon       {i\omega_{n}}
\def\igoi       {i\omega_{i}}
\def\igoj       {i\omega_{j}}
\def\gr         {\rho}
\def\gs         {\sigma}
\def\gS         {\Sigma}
\def\gSnko      {\Sigma_{\nk}(\go)}
\def\Skon       {\Sigma(\kk,\gon)}
\def\Snko       {\Sigma_{n\kk}(\go)}
\def\Snkoi      {\Sigma_{n\kk}(\goi)}
\def\gt         {\theta}
\def\gu         {\tau}
\def\gup         {\tau'}
\def\ek         {\gee_{\bf k}}
\def\ekpq       {\gee_{\bf {k+q}}}
\def\capo       {\right.\\ \left.}
\def\dn         {\downarrow}
\def\up         {\uparrow}
\def\rar        {\rightarrow}  
\def\la         {\langle}
\def\ra         {\rangle}
\def\dg         {\dagger}
\def\grG        {\gr_{nn'}(\kk,\qq,{\bf G})}
\def\grGo       {\gr_{nn'}(\kk,\qq,{\bf G_1})}
\def\grGts      {\gr_{nn'}^*(\kk,\qq,{\bf G_2})}
\def\uob        {\frac {1}{\gb}}
\def\uoZ        {\frac {1}{Z}}
\def\uot        {\frac {1}{2}}
\def\uop        {\frac {1}{\pi}}
\def\uon        {\frac {1}{N_{q}}}
\def\sumon      {\sum_{n=-\infty}^{+\infty}}
\def\sumoj      {\sum_{j=-\infty}^{+\infty}}
\renewcommand{\[}{\left[}
\renewcommand{\]}{\right]}
\renewcommand{\(}{\left(}
\renewcommand{\)}{\right)}
\def\MBPT       {many-body perturbation theory}
\def\MB         {many-body }
\def\GS         {ground state }
\def\GF         {Green's function }
\def\GFv        {Green's function, }
\def\GFs         {Green's functions }
\def\zpme         {zero point motion effect}

\def\efield	{ {\bf {\cal E}}}

\def\FF		    {{\bf F}}

\def\dk         {\frac{d\,{\bf k}}{\(2 \pi\)^3}}
\def\dku        {\frac{d\,{\bf k}_1}{\(2 \pi\)^3}}
\def\dq         {\frac{d\,{\bf q}}{\(2 \pi\)^3}}

\def\ecut       {\it E_{cut} }
\def\tpa        {{\it trans}--polyacetylene }
\def\Tpa        {{\it Trans}--polyacetylene }
\def\tpe        {polyethylene }

\def\Hscf       {H_{scf} }
\def\Vscf       {\widehat{V}_{scf} }
\def\Hzero      {\widehat{H}_0}
\def\Hone       {\widehat{H}_1}
\def\Htwo       {\widehat{H}_2}

\def\epc        {electron--phonon coupling }
\def\epcp        {electron--phonon coupling. }
\def\epcv        {electron--phonon coupling, }
\def\epi        {electron--phonon interaction }
\def\epip        {electron--phonon interaction.}
\def\epiv        {electron--phonon interaction,}
\def\ep         {electron--phonon }
\def\dw         {Debye--Waller }
\def\se         {self--energy }
\def\Se         {Self--energy }
\def\sep         {self--energy. }
\def\sev         {self--energy, }

\def\nk         {n{\bf k}}
\def\npkp       {n' \kk'}
\def\npk        {n' \kk}
\def\rnpk       {\mid  n'    \kk    \ra}
\def\lnpk       {\la   n'    \kk    \mid}
\def\rnk        {\mid  n     \kk    \ra}
\def\lnk        {\la   n     \kk    \mid}
\def\rnpkp      {\mid  n'    \kk'   \ra}
\def\lnpkp      {\la   n'    \kk'   \mid}
\def\rnpkpq     {\mid  n',{\bf k+q} \ra}
\def\npkmq      {n' {\bf k-q}}
\def\npkpq      {n' {\bf k+q}}

\def\uu		{{\bf u}_{Is}}
\def\uais	{u_{\ga,Is}}
\def\ubis	{u_{\gb,Is}}
\def\ubjsp	{u_{\gb,Js'}}

\def\tlim        {T going to zero limit}
\def\bql         {\hat{b}_{\qq \gl}}
\def\bmql        {b_{-\qq \gl}}
\def\bdql        {\hat{b}^{\dg}_{-\qq \gl}}
\def\bdpql       {\hat{b}^{\dg}_{\qq \gl}}
\def\gsq         {{\mid g^{\gql}_{n n' \kk} \mid}^2}
\def\Sf          {spectral function}
\def\Sfs         {spectral functions}
\def\qp          {quasiparticle}

\def\prl         {Phys. Rev. Lett. }
\def\prb         {Phys. Rev. B }
\def\pr          {Phys. Rev. }
\def\rpm         {Rev. Mod. Phys. }
\def\jpc         {J. Phys. Chem. }
\def\jp          {J. Phys. C }
\def\jcp         {J. Chem. Phys. }
\def\pss         {Phys. Stat. Sol. }

\def\self       {\epsfig{figure=Appendix/logo.ps,height=.5cm}\,}
\def\bverb      {\renewcommand{\baselinestretch}{1}\begin{verbatim}}
\def\everb  	{\end{verbatim}\renewcommand{\baselinestretch}{1.3}}

\def\ai         {{\it ab-initio} }
\def\ga         {\alpha}
\def\gb         {\beta}
\def\gC         {\Gamma}
\def\gD         {\Delta}
\def\gee        {\varepsilon}
\def\gl         {\lambda}
\def\gL         {\Lambda}
\def\go         {\omega}
\def\goql       {\omega_{\qq \gl}}
\def\gql        {\qq \gl}
\def\gS         {\Sigma}
\def\la         {\langle}
\def\ra         {\rangle}
\def\kk         {{\bf k}}
\def\qq         {{\bf q}}
\def\Gone       {{\bf G_{1}}}
\def\Gtwo       {{\bf G_{2}}}

\def\epc        {EP coupling }
\def\epcp       {EP coupling. }
\def\epi        {EP interaction }
\def\epip       {EP interaction. }
\def\ep         {EP }
\def\dw         {Debye--Waller }

\def\MB         {many body }
\def\se         {self-energy }
\def\sev        {self-energy, }
\def\sep        {self-energy. }
\def\Sf         {SF}
\def\Sfs        {SFs}
\def\qp         {quasiparticle}
\def\nk         {n\kk}
\def\npkp       {n' \kk'}
\def\npk        {n' \kk}
\def\rnk        {\mid  n     \kk    \ra}
\def\lnk        {\la   n     \kk    \mid}
\def\rnpkp      {\mid  n'    \kk'   \ra}
\def\lnpkp      {\la   n'    \kk'   \mid}
\def\uot        {\frac {1}{2}}

\def\Znk        {Z_{\nk}}
\def\zpme       {zero point motion effect}

\def\zbGN       {{\it zb}--GaN }
\def\ZbGN       {{\it Zb}--GaN }
\def\zbGNv       {{\it zb}--GaN, }
\def\zbGNp       {{\it zb}--GaN. }

\renewcommand{\[}{\left[}
\renewcommand{\]}{\right]}
\renewcommand{\(}{\left(}
\renewcommand{\)}{\right)}
\newcommand{\ltsim}{\protect\raisebox{-0.5ex}{$\:\stackrel{\textstyle <}{\sim}\:$}}

\title{Phonon-assisted shift-current from first-principles: application to AlN monolayer}

\newcommand{\bochum}{Research Center for Future Energy Materials and Systems and Interdisciplinary Centre for Advanced Materials Simulation, Faculty of Physics and Astronomy, Ruhr University Bochum, Universitätsstraße 150, D-44801 Bochum, Germany}
\newcommand{\cinam}{CNRS/Aix-Marseille Universit\'e, Centre Interdisciplinaire de Nanoscience de Marseille UMR 7325 Campus de Luminy, 13288 Marseille Cedex 9, France}
\newcommand{\holand}{MESA+ Institute for Nanotechnology, University of Twente,
7500 AE Enschede, The Netherlands}
\newcommand{\etsf}{European Theoretical Spectroscopy Facilities (ETSF)}
\newcommand{\lem}{LEM, UMR 104 CNRS-ONERA, Universit\'e Paris Saclay F-92322 Ch{\^a}tillon, France}
\newcommand{\torvergata}{University of Rome Tor Vergata, Rome, Italy}
\newcommand{\piim}{Universit\'e Aix-Marseille, Laboratoire de Physique des Interactions Ioniques et Moléculaires (PIIM), UMR CNRS 7345, F-13397 Marseille, France}
\newcommand{\texas}{Department of Physics, University of North Texas, Denton, TX 76203, USA}

\newcommand*{\PL}[1]{\textcolor{olive}{[PL: #1]}}
\graphicspath{{./Figures/}}

\author{Yuncheng Mao}
\affiliation{\bochum}
\affiliation{\cinam}
\author{Pierre Lechifflart}
\affiliation{\holand}
\author{Elena Cannuccia}
\affiliation{\piim}
\author{Claudio Attaccalite}
\email{claudio.attaccalite@univ-amu.fr}
\affiliation{\cinam}

\date\today
\begin{abstract}
In this work, we present a first-principles approach to investigate phonon-assisted shift current in extended systems. Our method combines the special displacement technique with real-time simulations based on the dynamical Berry phase. Merging these two frameworks enables direct access to non-linear optical responses at finite temperatures, while naturally incorporating phonon scattering and phonon-assisted electronic transitions into the time-dependent dynamics. This allows us to capture temperature-dependent features and dynamical lattice fluctuations that are essential for a realistic description of non-linear photocurrent generation.
As a representative example, we apply our approach to the shift-current response of a monolayer of hexagonal AlN (h-AlN). This system provides an ideal testing ground because it features an indirect band gap and lacks inversion symmetry. Moreover, h-AlN layers of different thicknesses have been fabricated and characterized in a variety of experimental settings, opening the concrete possibility of testing our predictions. Our results demonstrate the relevance of phonon-mediated processes and the importance of temperature in determining the magnitude and spectral features of the shift current in layered materials.
\end{abstract}

\maketitle

Thermal effects on electronic properties and optical responses have been known for decades\cite{giustino2017electron}. In recent years, several first-principles approaches have been developed to study bulk materials at finite temperature, based on the direct calculation of electron-phonon interactions,\cite{giustino2017electron} finite-difference atomic displacements,\cite{monserrat2018electron} or molecular dynamics\cite{garbuio2006ab}. These methods allow one to predict and reproduce band-gap renormalization,\cite{antonius2015dynamical,kawai2014electron} finite-temperature absorption,\cite{kawai2014electron,huang2021exciton,alvertis2020impact} and phonon replicas in absorption and emission spectra\cite{pillai,chen2020exciton}. However, most finite-temperature simulations have so far focused on linear optical responses, due to the intrinsic complexity of combining finite-temperature treatments with non-linear response calculations.

From an experimental perspective, non-linear optical measurements are commonly employed to investigate finite-temperature effects in biological systems\cite{hamal2021influence}. In condensed matter physics, coupling between lattice vibrations and non-linear response has been investigated in different systems and with different techniques: second-harmonic generation,\cite{PhysRevB.52.17158,khan2020extraordinary,mouchliadis2021probing} or four-wave mixing,\cite{8590813} the latter even exploited for temperature sensing\cite{8590813}.

Theoretically, temperature effects have been incorporated into linear-response calculations using finite-differences approaches\cite{huang2021exciton,alvertis2020impact} or by directly evaluating electron–phonon coupling.\cite{kawai2014electron,chen2020exciton} In contrast, \RED{only a few studies have addressed non-linear response at finite temperature in bulk materials\cite{anna}}, while for molecular systems molecular dynamics\cite{watanabe2010half}, or on multiconfigurational Ehrenfest dynamics\cite{chen2021simulation} have been used to obtain non-linear response functions at finite temperature.

Here, to overcome these limitations, we present a first-principles framework for computing non-linear optical properties at finite temperature. Our approach combines real-time simulations for the evaluation of non-linear response functions with the special displacement technique to take into account coupling with phonon modes and temperature effects. In particular, we applied the proposed methodology to the shift-current response.

In recent years, there has been a growing interest in understanding the role of phonons and electron–phonon coupling in the bulk photovoltaic effect (BPVE). In particular, recent studies have investigated how quantum geometry\cite{hu2024optical} and shift current\cite{morimoto2025role,akay2026phonon,wang2026} are modified by electron–phonon interactions. Another study used a first-principles density matrix approach to incorporate relaxation effects mediated by both phonons and photons\cite{yu2026} in the shift current response. Meanwhile, an experimental group reported the first observation of phonon-assisted shift current in twisted bilayer MoS$_2$\cite{huang2025phonon}.

In this work, we used h-AlN monolayer as prototype material to study phonon-assisted shift current. This material has been recently synthesized in its hexagonal form\cite{tsipas2013evidence}, and it has been shown to go from direct to indirect with decreasing number of layers\cite{bacaksiz2015hexagonal}. Phonons and electron-phonon coupling in h-AlN have been recently investigated using Raman\cite{jin2016deep} and infrared spectroscopy\cite{prokofyeva2001vibrational}.  
Monolayer h-AlN is an ideal testing case for our study because it has an indirect band gap, lacks inversion symmetry, and h-AlN structures of different thicknesses have been fabricated and characterized in a variety of experimental settings.
\begin{figure*}[th]
    \centering
	\includegraphics[width=1.00\textwidth]{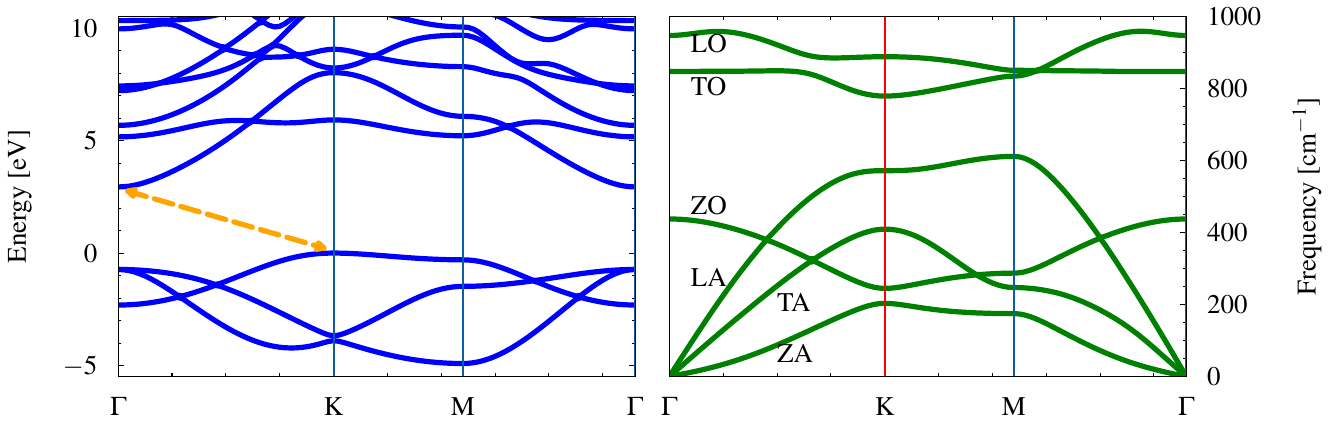}
        \caption{\label{fig:aln_bands} Left panel: Kohn-Sham band structure along the high symmetry lines. Right panel: phonon dispersion along the same path. In the left panel the indirect gap is highlighted by an orange arrow, while the direct gap is situated at the $\mathbf \Gamma$ point.}
\end{figure*}

To demonstrate this approach, we detail below the integration of real-time Berry phase dynamics with the special displacement method, and subsequently apply the framework to evaluate the phonon-assisted shift current in monolayer h-AlN


We briefly present the definition of the SC response and the real-time approach used to calculate it, as well as the procedures to extract nonlinear conductivity coefficients from the real-time current. 
When a material is irradiated by light, a current is produced. In the perturbative regime, this current can be expressed as a power-series expansion of the total field strength:
\begin{equation}
	J(\omega)=\sigma^{(1)}  E(\omega) + \sum_{\boldsymbol{\omega}} \sigma^{(2)} E(\omega_1)E(\omega_2) + O(E^3),     
\end{equation}
where the summation on $\boldsymbol{\omega}$ indicates the sum over all distinct $\omega_1,\omega_2$ pairs.  
With $\omega_1, \omega_2 \neq 0$, the lowest order that produces DC current is the second order when $\omega = \omega_1 + \omega_2 = 0$.~\cite{PhysRevB.61.5337} In case of a monochromatic field,
\begin{equation}
J^{(2)}(\omega=0)= 2\sigma^{(2)}(0;\omega_1,-\omega_1) E(\omega_1)E(-\omega_1).
\end{equation}
We simulate a system under the influence of an external, time-dependent electric field by solving an effective Schr\"odinger equation:
\be
i\hbar  \frac{d}{dt}| v_{\kk,m} \rangle = \left(\hat H_\kk + i {\bf \cal{E}}\cdot \partial_\kk  + \hat \Gamma_\kk \right)| v_{\kk,m} (t) \rangle, \label{eom}
\ee
where $\hat H_\kk$ is the effective Hamiltonian, $| v_{\kk,m} (t) \rangle$ are the time-dependent valence states and the dipole operator has been replaced with $\rr = i \partial_\kk$ and the k-derivative has been performed using a covariant approach\cite{souza_prb,attaccalite2013nonlinear}. The term $\hat \Gamma_\kk$ is a dephasing term defined as:
\be
\hat \Gamma^{\kk}=\frac{i}{\tau} \left \{ \hat \rho_{\kk} -\hat \rho^0_{\kk} \right \}
\label{eq:dephasing}
\ee
where $\rho_{\kk}$ and $\rho^0_{\kk}$ are respectively the density matrix at time $t$ and at equilibrium, and $\tau$~ is a dephasing time. 
In the following, we will omit the time-dependence of $| v_{\kk,m} (t) \rangle$ in order to simplify the notation. From the real-time dynamics of the valence states we calculate the current as:
\be
J=\frac{f}{(2 \pi)^3} \sum_{n=1}^M  \langle v_{\kk n}|\hat H_\kk \partial_\kk | v_{\kk n}\rangle + c.c.
\label{eq:current}
\ee
The derivative in $\kk$ space is evaluated by means of finite-difference covariant formulation\cite{souza_prb,attaccalite2013nonlinear}.  
To obtain the second order conductivity corresponding to the SC,  we perturb the system with a sinusoidal electric field $E(t)=E_0 \theta(t-t_0)sin(\omega_0 t)$, where $E_0,\omega_0$ are the strength and frequency of the applied electric field, respectively, and $\theta$ is the Heaviside step function. Further, we add a term, Eq.~\ref{eq:dephasing}, to the Hamiltonian to simulate an electronic dephasing and to wash out the  eigenmodes that are excited when the field is switched-on at $t_0$ \cite{attaccalite2013nonlinear}. After transient effects have decayed, the final current can be written as a Fourier series expansion: $J(t) = \sum_{j=-N}^N c_j e^{\mathrm{i}j \omega_0 t}$. The coefficients $c_j$ can be obtained by a discrete Fourier transform by sampling $J(t)$ on a period $2\pi/\omega_0$~\cite{paleari2025yambopy}. The components of the $n$-order conductivity tensor are directly related to the $c_j$ coefficients through powers of the applied electric field. 
In the shift-current case, we extract the conductivity corresponding to the coefficient $c_0 = 2\sigma^{(2)}(0;\omega_0,-\omega_0) E_0^2 + O(E_0^4)$\cite{mao2025shift}.  This procedure is the analogue of that used to extract the nonlinear susceptibilities from the polarization density that has been detailed in Refs.~\onlinecite{pionteck2025sum,attaccalite2013nonlinear}.


In order to include coupling with lattice vibrations in our real-time dynamics we use the Special Displacement Method (SDM) as presented in Ref.~\onlinecite{sdm}. The SDM provides a framework to compute temperature-dependent properties by mapping the stochastic sampling of the nuclear vibrational ensemble onto a single, supercell calculation.\\
The central idea of the SDM is to approximate the expectation value of an electronic observable $\hat{O}$ at a finite temperature $T$ using a single configuration of the supercell. This is expressed as:

\begin{equation}
\langle \hat{O} \rangle_T \approx O(\mathbf{R}_T).
\end{equation}

Here, $\mathbf{R}_T$ is the special displacement, a single, specific configuration of the atomic coordinates that captures the quantum-statistical distribution of the nuclear positions at temperature $T$. In the SDM we suppose that the observable we are interested in, for example in our case the shift current, can be approximated with a quadratic form around the equilibrium configuration:
\begin{equation}
\hat{O}(R) = \hat{O}_0 + \left.\partial_R \hat{O}\right|_{R_0} \Delta R + \frac{1}{2}\left.\partial_R^2 \hat{O}\right|_{R_0} (\Delta R)^2 + \mathcal{O}(\Delta R^3) \label{O2}
\end{equation}
where the displacement $\Delta R$ has to be considered along a phonon mode. In the normal mode basis all phonon modes are independent harmonic oscillators and therefore we easily average a quadratic operator on their probability distribution that is a Gaussian both at zero and finite temperature. Now if we suppose that the opertor we are interested in  has a quadratic form as in Eq.~\ref{O2}, it can be shown that a single calculation yields the exact thermal average.\\
This special configuration $\mathbf{R}_T$ is constructed in the basis of the vibrational normal modes of the system. For a supercell containing $N$ atoms, the displacement of atom $s$ in the supercell from its equilibrium position $\boldsymbol{\tau}_s$ is given by:

\begin{equation}
\mathbf{R}_T(s) = \boldsymbol{\tau}_s + \frac{1}{\sqrt{N}} \sum_{\mathbf{q}, \nu} \Delta_{\mathbf{q}\nu}(T) \, \mathbf{e}_{s \nu}(\mathbf{q}) \, e^{i \mathbf{q} \cdot \mathbf{r}_s}.
	\label{sdm_r}
\end{equation}

The crucial component is the {special displacement amplitude $\Delta_{\mathbf{q}\nu}(T)$, which is a complex number defined as:

\begin{equation}
	\Delta_{\mathbf{q}\nu}(T) = X_{\mathbf{q}\nu} \sqrt{\frac{\hbar}{2 M_s \omega_{\mathbf{q}\nu}} \left ( 2 n_{\mathbf{q} \nu} (T) + 1 \right)} 
\end{equation}

In this equation: $\omega_{\mathbf{q}\nu}$ and $\mathbf{e}_{s \nu}(\mathbf{q})$ are the frequency and eigenvector of the phonon mode with wavevector $\mathbf{q}$ and branch $\nu$; $M_s$ is the atomic mass of atom $s$; $ n_{\mathbf{q} \nu} (T)$ is the Bose-Einstein distribution function and  $X_{\mathbf{q}\nu}$ are $\pm 1$ coefficients. In Ref.~\onlinecite{sdm} it has been shown that is possible to speed the thermodynamic limit with the supercell by optimizing the displacemts signs and the phase of the phonon eigenvectors in order to have a continous phase change in the Brillouin zone. We adopted all these strategies in generating of special displacement supercells following the procedure described in Ref.~\onlinecite{sdm}.
\begin{figure}
    \centering
	\includegraphics[width=0.48\textwidth]{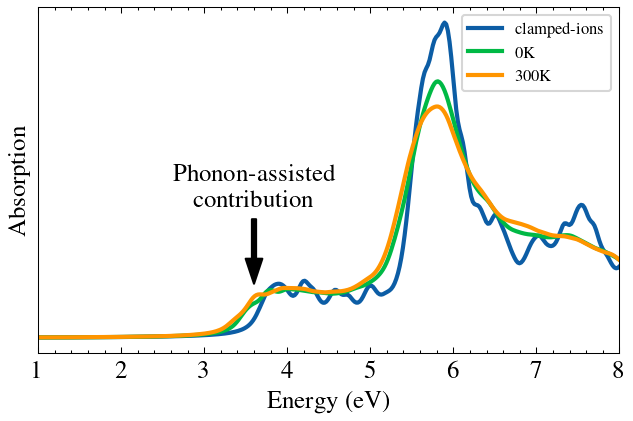}
        \caption{\label{fig:aln_eps} Optical absorption of AlN monoayer calculated at clamped-ions and with SDM at 0 and 300 Kelvin using an $8\times8$ supercell with SDM.}
\end{figure}

The primary advantage of the SDM is its practical implementation. The atomic coordinates $\mathbf{R}_T(s)$ define distorted configuration in supercells. This supercell can be used as the direct input for our real-time dynamics.\\
The method is formally exact for observables that are quadratic in the atomic displacements. For non-linear observables, the SDM remains a highly accurate and efficient approximation, as it is designed to capture the most significant anharmonic effects through the specific construction of $\Delta_{\mathbf{q}\nu}(T)$. 
In actual calculations, the complex displacement vector, Eq.\ref{sdm_r}, is mapped in a real-space displacement in a supercell compatible with the $\qq$-grid. Then the number of $\qq$-points and the corresponding size of the supercells, is increased in such a way to converge to integration in $\qq$-space.

In order to simulate an isolated h-AlN monolayer we used a supercell with a size along z-direction of $27.5~a.u.$ and a lattice constant of $3.12$ \AA\, in the plane. The ground state of h-AlN monolayer was studied by means of Quantum-Espresso (QE) code\cite{pwscf} using DOJO pseudo-potentials\cite{PseudoDojo}, LDA exchange correlation functional, a cutoff for the wave-functions of 90~Ry and a $18 \times 18 \times 1$ k-grid. Phonons were calculated on a $9 \times 9 \times 1$ q-grid  and then interpolated using the force constant method\cite{pwscf}.
    Linear and non-linear optical responses were calculated using the Lumen branch\cite{lumen} of the Yambo code\cite{yambo}. For the linear response we used the standard linear response theory\cite{yambo}, while for the shift-current we used the real-time approach described above. For both response functions we used bands from 2 to 7 and $24 \times 24 \times 1$ k-grid. When we build the supercell for the SDM we rescale k-grid in such a way to keep the density of k-points constant in the Brillouin Zone.

\begin{figure}
    \centering
	\includegraphics[width=0.5\textwidth]{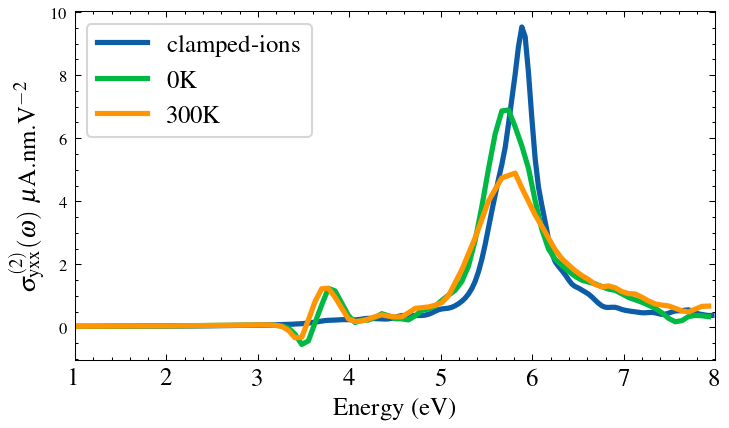}
        \caption{\label{fig:aln_sigma_yxx} Shift current response calculated at clamped ions and including quantum ions at 0 and 300 Kelvin using an $8\times8$ supercell with SDM.}
\end{figure}

In our simulations, we included a small amount of dephasing in the dynamics,(see Eq.~\ref{eq:dephasing}), of $0.05~eV$, in order to simulate intrinsic electronic dephasing that is not due to coupling with phonons, and to eliminate the eigenmodes excited by the sudden turn-on of the field\cite{pionteck2025sum}.
For the simulation with SDM we used a  $8 \times 8$ supercell and we verified that this size is sufficient to converge both linear and non-linear response functions.\\
In Fig.~\ref{fig:aln_bands} we report the electron and phonon band structures for monolayer h-AlN. For the electronic part, left panel of Fig.~\ref{fig:aln_bands},  we found an indirect band gap between $\mathbf \Gamma - \mathbf K$ of $2.94$ eV and a direct one at $\mathbf \Gamma$ of $3.67$~eV, in agreement  with previous publications. In the right panel of Fig.~\ref{fig:aln_bands}, the phonon dispersion is reported. In particular we draw in red the line corresponding to the phonon momentum that connect the valence band maximum and conduction band minimum of the indirect band gap. In general we expect that phonon modes between  $\mathbf \Gamma - \mathbf K$ will be the one responsible for the phonon-assisted processes, because their momentum is compatible with indirect transitions between valence and conductions bands.\\

In Fig.~\ref{fig:aln_eps}, we report the optical absorption calculated with clamped ions and using the SDM method at 0 and 300~K with an $8 \times 8$ supercell. We find that introducing the coupling with phonon modes reduces the intensity of the main peak by about 16\%. This reduction is further enhanced when the temperature increases from 0 to 300~K, and a small red-shift of the main peak is present due to the gap reduction. Across the rest of the spectrum, the sharp features resulting from the small smearing used with clamped ions are washed out by the damping induced by phonon modes. In addition to the intensity reduction, non-zero absorption appears below the optical gap (indicated by the arrow in Fig.~\ref{fig:aln_eps}) due to phonon coupling, which enables otherwise forbidden indirect transitions. We estimate that the coupling with phonons lowers the absorption threshold by about 0.3~eV. While this is not equivalent to the 0.7~eV difference between the direct and indirect band gaps, one might expect lower-energy transitions to become increasingly weak and therefore invisible on this scale\cite{giustino_silicon}.\\
Then, we move to the shift current (SC) response $\sigma^{(2)}_{yxx}(\omega)$, calculated using the real-time approach previously presented.  We found that temperature effects reduce the maximum SC response by more than 27\%, which is much more pronounced than the reduction seen in the linear response.  This behavior is expected because, in the case of SC, electron-phonon coupling can generate contributions with negative signs that are otherwise absent in the optical absorption. We also observed a shift in the position of the maximum response, which can be attributed to band gap reduction; however, it is important to keep in mind that for non-linear response functions, a gap reduction does not necessarily correspond to a rigid shift of the spectrum.
Then, precisely at the points corresponding to the indirect transitions mediated by phonons that we observed in the linear response, a new small peak appears in the shift current (SC) response, exhibiting alternating signs. This is a very interesting feature that demonstrates how electron-phonon coupling not only suppresses the SC response but, in materials with indirect band gaps, can also generate a new shift current response mediated by indirect phonons where it was previously absent. \\
In the next section, in order to gain deeper insight into this phonon-assisted response, we directly analyze the response to single phonon modes at a fixed momentum.
To clarify the microscopic mechanisms, we analyze the specific phonon modes responsible for these spectral features. This is not an easy task, as different modes can contribute to the formation or modification of a single peak in the spectrum at different moments. We consider two fixed phonon wavevectors and analyze how the different modes modify the response. Specifically, we examine the momentum $\mathbf{q}=0$ and the momentum $\mathbf{q}=\mathbf{K}$, the latter of which connects the indirect band gap. We expect these $\mathbf{q}$-points to be representative of the indirect transitions mediated by all phonons spanning between these two extremes. Figure~\ref{fig:shift1} shows the distortion of the shift-current response for different phonon modes at $\kk$ and $\mathbf{\Gamma}$ point. For each mode we displace the atoms according to Eq.~\ref{sdm_r}, but considering each mode separately. We found that at $\mathbf K$ point the modes responsible for the phonon-assisted response are the ZO, LA and TO, while the other modes ZA,TA and LO only affects the intensity of the main peak. At $\mathbf{\Gamma}$ only three modes are present because the acoustic ones have zero coupling. We found again that also at $\Gamma$ the modes responsible for the formation of the phonon-assisted peak are the ZO and TO modes, while the LO mode merely attenuates the main peak. It is interesting to note that coupling with the TO mode at $\mathbf {\Gamma}$ yields a negative signal between 3 and 4 eV. Clearly the coupling with only this two momenta are not enough to explain all the spectrum generated in the $8 \times 8$ supercell where we expect the interference of the different modes and momentums is at the origin of the final spectrum.  However, these findings clearly indicate that while specific optical and acoustic branches (such as ZO, TO and LA) are actively driving the microscopic mechanism behind the phonon-assisted sub-gap photocurrent, other modes primarily dictate the overall scaling and damping of the primary response. This selective mode coupling underlines the highly non-trivial quantum mechanical interference occurring in non-linear responses.

\begin{figure*}
    \centering
        \includegraphics{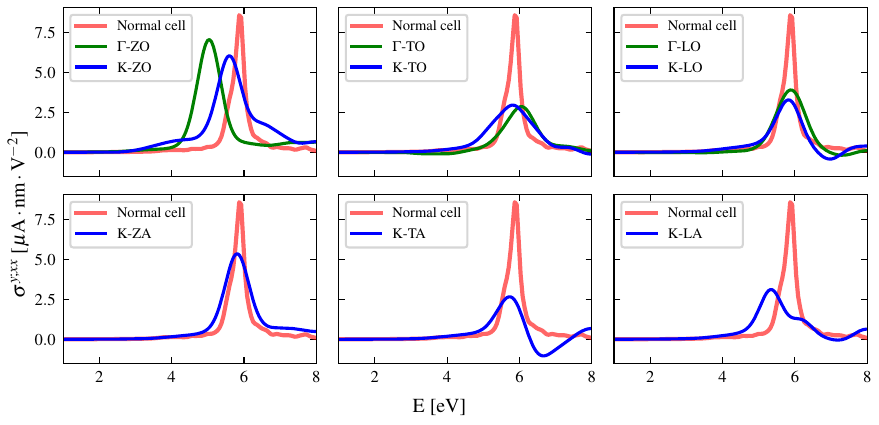}
	\caption{\label{fig:shift1} Distortion of the shift-current response $\sigma_{y;xx}$ for different phonon modes at $\mathbf \Gamma$ and $\mathbf{K}$ points. Phonon nomenclature follow the $\mathbf \Gamma$ point, see right panel of Fig.~\ref{fig:aln_bands}.}
\end{figure*}



In this work, we have presented a novel first-principles framework that merges the Special Displacement Method (SDM)\cite{sdm} with real-time simulations based on the dynamical Berry phase\cite{pionteck2025sum}. This combined approach successfully circumvents the historical bottleneck of calculating non-linear optical responses at finite temperatures, allowing for the direct integration of dynamical lattice fluctuations and phonon-assisted electronic transitions into time-dependent quantum dynamics. 
We apply this methodology to a prototype $h$-AlN monolayer, which is an ideal system characterized by an indirect band gap and broken inversion symmetry. First, we observed that incorporating electron-phonon interactions results in a substantial suppression of the main shift current (SC) response peak. This reduction is notably more pronounced than the attenuation observed in the linear absorption spectrum. We attribute this enhanced sensitivity to the fact that electron-phonon coupling in the non-linear response can generate quantum mechanical contributions with alternating or negative signs, which actively interfere with and suppress the peak response—a mechanism intrinsically absent in purely linear absorption processes. Most significantly, our framework successfully captured a purely phonon-assisted shift current. Below the direct optical gap, precisely at energies corresponding to the indirect transitions observed in linear absorption, a new, sub-gap peak characterized by alternating signs emerges in the SC spectrum. 
This finding demonstrates that while phonon scattering generally suppresses the overall magnitude of the bulk photovoltaic effect (BPVE) at higher energies,  it plays a constructive role in materials with indirect band gaps by opening up novel, momentum-conserving conduction channels mediated by indirect phonons. \\
In conclusion, we think that the combinantion of the SDM combined with real-time Berry phase dynamics offers a scalable roadmap to explore temperature-dependent non-linear phenomena in a wider class of materials. 
\begin{acknowledgements}
	C.A. and Y.M. acknowledge B. Demoulin and  A. Saul for the management of the computer cluster \emph{Claudia}. C.A. and Y.M. acknowledge ANR project COLIBRI No. ANR-22-CE30-0027. This work is supported by the Horizon Europe research and innovation program of the European Union under the Marie Sklodowska-Curie grant agreement 101118915 (TIMES).  C.A. acknowledges support from AMUtech through the INDIGENA grant (AMX-22-RE-AB-208). 
\end{acknowledgements}


\bibliography{nloptics,nonlinear_finiteT}
\bibliographystyle{apsrev4-1}
\end{document}